\documentclass[11pt]{article}

\usepackage{mathpple}
\usepackage{graphicx}
\usepackage{url}
\usepackage{paralist}

\usepackage{geometry}
\begin{document}

\noindent
\textbf{Preprint of:}\\
T. A. Nieminen\\
``Project-based assessment for graduate coursework in physics''\\
in R. Sang and J. Dobson (eds),
\textit{Australian Institute of Physics (AIP)
  17th National Congress 2006: Refereed Papers},
Australian Institute of Physics, 2006 (CD-ROM, unpaginated).

\hrulefill

\begin{center}

{\huge
\textbf{Project-based assessment for graduate coursework in physics}}

\vspace{2mm}

\textit{T. A. Nieminen}\\
School of Physical Sciences, The University of Queensland, Australia

\subsection*{Abstract}

\end{center}

\begin{quote}
Project-based assessment, in the form of take-home exams, was trialed in
an honours/masters level electromagnetic theory course. This assessment
formed an integral part of the learning experience of the students, and
students felt that this was effective method of learning.
\end{quote}

\section*{Introduction}

Graduate coursework in physics---the coursework component in honours and
masters courses---is intended in large part to prepare students for
postgraduate research and the professional practice of physics.
Assessment often depends largely, or even entirely, on examinations of
the traditional variety---proofs and problems, to be completed in
perhaps two or three hours. A necessary result of this format is that
the problems must be simple enough for able students to complete
satisfactorily in the time available. As I have rarely encountered the
need for such proofs or solutions to such simple problems in the
professional practice of physics, and in any case, these are the type of
things which can be easily looked up textbooks, I wondered as to how
well such assessment methods meet the intended requirements of the
courses.

Furthermore, the capability of students to memorise required proofs and
solutions until the end of the examination, and then to promptly forget
a large portion of this, suggests that long-term learning is not
enhanced by traditional examinations.

Considering the above points, I entertained the possibility of
alternative forms of assessment. Observations on the lack of correlation
between examination results and my perceptions of student understanding
provided further incentive.

Accordingly, I replaced traditional examinations in an
honours/masters-level course in electromagnetic theory with
project-based take-home exams. Students would complete one question in
approximately one week, and could make use of the tools available to
professional physicists: the research literature, libraries, and
colleagues. Questions were chosen to require an estimated two days of
work.

I discuss the implementation of project-based assessment, and the results
of two years of testing. Finally, I consider some future possibilities.

\section*{Implementation}

Two project-based take-home exams were conducted in the discussed
course in each of 2005 and 2006. In both cases, in the first instance,
the topic was electrostatics, and all students attempted the same
question. In the second instance, the topic was electromagnetic
waves, and students could choose one question from a total of five.
Students were usually given five or six days in which to attempt the
question; this time included one weekend. Questions were designed
with the intent that two student-days would be required to
complete the tasks. The topics had been covered beforehand in
lectures.

For example, the first project-based exam in 2005 was:
\vspace{-6mm}
\begin{quotation}
\noindent
A colleague approaches you for advice. She is planning an
experiment in which a cubic conductor will be charged to
a potential. To simplify the analysis, she hopes that it will
be possible to assume that the field is a Coulomb field (ie like
the field due to a point charge). How closely does the field of
a conducting cube resemble that of a conducting sphere?

\noindent
You might wish to consider some of the following questions:
\begin{enumerate}
\item How far away from the cube do you need to be before the field
closely resembles the field due to a sphere?
\item Is the ``equivalent sphere'' at the same potential as the cube?
\item What is the field due to a conducting cube?
\item What about the effect of the ground? Assume that the ground
can be treated as a perfectly conducting infinite plane.
\end{enumerate}
\end{quotation}

The range of questions for the second exam is illustrated by the
2006 selection:
\vspace{-12mm}
\begin{quotation}
~
\begin{enumerate}
\item
Investigate the transmission and reflection of
long pulses by an anti-reflection dielectric layer
(or some other interesting type of layer(s)).

You may wish to perform the calculation for a periodic sequence of
pulses.
\item
What is the extinction cross-section of a sphere?
Calculate the cross-section as a function of radius for
a variety of types of sphere.

Under what conditions might experimental measurements reproduce
such curves?
\item
What is the smallest resonator that can be made
from a circular loop of optical
fibre that has a $Q$ close to that of a large circular loop of optical
fibre?
\item
Design a passive antenna array, with one driven element, or an active
array, with all elements driven, for broad-band
directivity. Calculate how the radiation pattern changes as a
function of frequency.
\item
Is it possible to build a Bragg mirror that is large enough so that
its structure can be clearly seen and can be used with a laser pointer
for demonstration purposes?

Calculate or estimate the maximum thickness of each layer.
\end{enumerate}
\end{quotation}

It was intended that the question in the first exam could be
approached either theoretically or computationally, while the
second exam included both questions suited to a theoretical
approach (eg Q3 above), questions suited to a computational
approach (eg Q2), and those amenable to either.

Students were required to hand in a written report, and to
make an oral presentation in class. There were no specific requirements
for either the written report or oral presentation, other than
clear communication of methods and results.

The two project-based exams together constituted approximately
40\% of the total assessment for the course, with the remaining
assessment consisting of assignments and minor oral or written
in-class exams.

\section*{Results}

Students were, in almost all cases, able to complete the
exams in the time allowed. The time spent by the students on
the questions was greater than expected, with three days being
typical.

The variety of methods employed by the students was surprising;
for the first exam in 2005, where all students attempted the same
question, there were five completely distinct strategies used
by the 13 students. Similarly, for the second exam in each year,
each of which had five questions to choose from, every question
was attempted by at least one student.
Students frequently chose
computational methods of solution.
While students did make use of the research literature, this was
to a lesser extent than expected. This is likely to be a result
of many of the computational methods being relatively
straightforward.

Student performance varied from good through to superb.
The quality of presentations noticeably improved (students
also gave oral presentations for a minor assignment and for the
major assignment). Students were able to demonstrate an
understanding of the problem and the related physics and
mathematical or computational methods beyond merely
demonstrating the ability to provide a numerical answer or
proof.

Written reports varied from seven pages through to over 30, with
15 pages being typical, not including attached listings of computer code.
Thus, these assessment tasks also provide a significant degree of
practice in written scientific communication.

Some students made extensive use of the research literature, while
others made little or no use of literature other than textbooks.
In part, this depended on the nature of the question chosen,
but also reflected individual approaches to the problems.

\subsection*{Student feedback}

Students generally found that the problems were difficult
and required a significant amount of work. Despite this, they
also found the experience enjoyable. Students reported that
they learned more from the project-based assessment than any
other assessment. Several students recommended that the
number of project-based take-home exams be increased, so
as to cover a larger amount of the course content.
Student feedback included both unsolicited and solicited feedback,
primarily verbal. 

\section*{Discussion}

Overall, I believe that the project-based assessment described
above was successful. Rather than simply testing knowledge of
and ability to apply the course material, the take-home exams
provided an intensive learning experience for the students.
This learning included not only the relevant portions of the
core course content, but also useful skills such as how to make
use of the research literature and oral and written scientific
communication.

Based on the results of this trial, the number of such
project-based take-home exams will be increased, possibly
to five, and the entire course content will be covered.
Prior to the first assessment task for a semester, the assessment method
will be discussed with the class, and guidelines for the
written report and oral exam will be given. The discussion
will include ethical standards and how to make use of the
research literature. While a formal discussion of these
issues may not be necessary for honours or masters level
students, I believe that it could still benefit them.
Such a discussion would be essential if the course
were an undergraduate course.

In 2005 and 2006, the other major component of the assessment
was an assignment on a relevant topic of the student's choice.
The above plan to increase the number of project-based take-home
exams will essentially result in the elimination of the assignment.
This will have the advantages of spreading the work more
evenly over the entire semester---most students leave the bulk of
the work on the assignment until shortly before the due date---and
assessing a larger part of the course material. Although
the assignment allowed students to investigate one aspect of the
course in depth, this was at the cost of breadth. It may be
of value to retain an assignment as an optional piece of assessment,
perhaps to replace one or two take-home exams that cover similar
material, for students who wish to attempt publication-quality
work---approximately 20\% of the students have produced work
of this standard in 2005 and 2006 (for example, Pfeifer and
Nieminen~2006), and it would be a pity to deny students this
opportunity.

The time required for the oral presentations is significant,
and when the number of such exams is increased, a significant
portion of the total class time during the semester will
be occupied by such presentations. However, I feel it is
important to retain the oral presentations, since scientific
communication is an important learning objective for
courses intended for the training of professional physicists.
The imposition of strict time limits for presentations is
necessary---when students were given a recommended time of
15 minutes for presentations, while some students kept their
presentations within this time frame, others gave 40 minute
presentations. With 12 students giving presentations of 15 minutes
each for four take-home exams, approximately $1/4$ of a
4 contact hour per week course will consist of presentations.
If such take-home exams replace a traditional final exam,
there is no need for drill problems to prepare students
for the final exam, which often occupies a large part of
tutorial time, this time should be available.

A number of issues, which include potential problems,
deserve further consideration. 

\subsection*{Large classes}

Large classes may not allow enough time for students to
give oral presentations for every exam.
Some possible solutions include students giving only one
or two oral presentations over the semester, eliminating
oral presentations entirely (but see the next item below),
or group-work. If students were to work in groups, then
one student from each group could give the presentation
on behalf of the group, with a different student presenting
for each exam. Group-work would also reduce the time required to
read the written reports if the group submits a single report.
This, of course, would leave one with the problem of
fairly determining grades for individual students.

\subsection*{Plagiarism}

The oral presentation, with the opportunity to question
the student, is an important part of determining the level
of understanding that the student has. As a result, it is
useful for the detection of plagiarism---if the student has
not done the work, they are less likely to understand it.
It is important to clearly communicate the ethical
standards expected of the students, the extent to
which they can make use of the work of others, and how to
properly acknowledge such use of other's work.

\subsection*{Workload}

The effort demanded of students for a take-home exam is significant,
and potential clashes with other major demands on student time
should be watched for.

\subsection*{Instant answers}

For the first exam each year, where all students attempted the
same question, at about the time when students were completing the
exam after almost a week of effort, it became apparent that it was
possible to cover most elements of the problem in under five minutes.
For example, with the cube problem given above, the symmetry required
of the potential means that the first non-zero multipole moments
beyond the monopole moment are the $2^5$ terms (only the odd-$n$
$2^n$ terms can be non-zero, and the $2^3$ terms must be zero
because they cannot give cubic symmetry), and hence all non-monopole
contributions to the potential drop off at least as fast as $r^5$.
With this, one can very quickly provide a reasonable answer to
the question as posed. 

\subsection*{Impossible questions}

One interesting possibility is to set questions that are
difficult to the point of impossibility. The paths by which
the student attempts to approach the answer, and the points
at which these are abandoned may well demonstrate the student's
understanding better than a straightforward successful solution.
It may well be wise to warn students beforehand of the possibility
of impossibility to avoid excess student anguish and stress.

\subsection*{Experimental answers}

The inclusion of problems that can include or require experimental
measurements can allow a close integration of theory and experiment.
This could be a useful strategy for adding a small experimental
component to a theoretical course.

\subsection*{Summary}

Project-based take-home exams were a successful learning and
assessment method, and were popular with the students.
Students reported that they learned more from them than
from other types of assessment, and recommended that the
number of such take-home exams be increased. This recommendation
will be adopted.

\section*{Refererence}

Pfeifer, R.N.C and Nieminen, T.A. (2006). Visualization of \v{C}erenkov
radiation and the f{i}elds of a moving charge.
\textit{European Journal of Physics} \textbf{27}, 521-29.

\end{document}